\documentstyle[prl,aps,twocolumn,epsf]{revtex}
\def\break#1{\pagebreak \vspace*{#1}}
\begin{document}

\draft

\title{Superoscillations and transplanckian frequencies}

\author{H. Rosu\cite{byline}
}
\address{ 
{Instituto de F\'{\i}sica de la Universidad de Guanajuato, Apdo Postal
E-143, Le\'on, Gto, M\'exico}\\
}

\maketitle
\widetext

\begin{abstract}
{\bf Summary.} - Superoscillations can explain the arbitrarily
high frequencies' paradox in black hole radiation.

\end{abstract}

\pacs{PACS 04.60+n - Quantum theory of gravitation
\hfill LAA number: gr-qc/9606070
}

\narrowtext

If one traces, in the usual quasi-geometrical derivation, the origin of the
thermal radiation from a solar mass sized black hole \underline{a second}
after the formation of the black hole, its origin lies in the incoming vacuum
frequencies of the order $e^{10^4}$ (units are not important), implying scales
of orders by far larger than the mass of the universe itself. This is
almost a quotation from Unruh's work and represents the paradox of
transplanckian frequencies, which
is one of the most important unsolved issues in black hole radiation, and
in fact, might be an indication of a serious drawback of the employed
formalism. It has been first discussed by Jacobson \cite{jac} and by
Unruh \cite{u} in a hydrodynamical model.
The conclusion of Unruh after a quite detailed analysis of
two-dimensional `dumb' holes is that there is
an adiabatical adjustment of high frequency phenomena to low frequency physics
close to the horizon (no high-frequency quantum hair). Since
short distance cutoffs (at the Planck scale) are not affecting Hawking's
thermal radiation, the common trend in the literature is to get rid of
the transplanckian reservoir, e.g., by finding physical models (borrowed
from other fields of physics) for the cutoffs \cite{tj}.

Another way to get out from the transplanckian absurdity is to think of
a material support of those extremely high frequencies, and if possible,
of a physical entity and phenomenon related to black hole physics.
Let me spell now my answer of principles
to solve the transplanckian paradox, which
I associate to {\em superoscillations},
a remarkable phenomenon put forth by Aharonov \cite{aapv}.
Recall the underlined second in the first paragraph. This
would correspond to a band-limited thermal signal of 1 Hz. Apparently,
it is not well-known that band-limited functions can oscillate
for arbitrarily long intervals arbitrarily faster than the highest
frequency they contain. These are the superoscillations. One can obtain them
through a special two-parameter integral representation of band-limited
functions (BLFs), which, in general, changes a  BLF to a corresponding
two-parameter superoscillating function (SOF). As intuitively
presented by Berry, the special integral representation of BLFs
contains a Gaussian-like factor, which, in the zero-width limit acts like
a $\delta$-function and projects out SOFs.
These SOFs appear in a
subtle way in the `quantum time-translation machine' concept of
Aharonov, Anandan, Popescu, and Vaidman \cite{aapv}, but the present note
has its origin
\break{0.72 in}
in my repeated reading of Professor Berry's works \cite{b}.
Among the physical examples of superoscillations given by Berry,
that of special interest for my argument here
is the two-dimensional case of evanescent waves which can be expressed as
the singular limit
of an angular superposition of real plane waves. The main point now
is that the in-out formalism does not include the quantum hair, i.e.,
the evanescent waves at the horizon.
Any evanescent hair has the property that
in the direction perpendicular to the exponentially decaying one (thus along
the horizon), oscillates
faster than the free-space wavenumber, forming a
{\em superoscillating reservoir}.
One can construct angular
superpositions of `plane' waves for a Gaussian
wave packet in the black hole background and show readily, just following
Berry, that its singular limit is a superoscillatory function \cite{rosh}.
Also, enclosed vortex configurations are important for the
energy-momentum transport by means of evanescent modes.

Moreover, superoscillations may be related to a well-known problem in
signal processing, that of sampling a function faster than the Nyquist rate
(Berry is quoting Daubechies), and also to quantum billiards, which would
be very interesting to study in the black hole context.
Berry gave an explicit example of reproducing Beethoven's ninth symphony
(a signal spreading over 4000 s) with a signal bandlimited to 1 Hz
by means of superoscillations. As found by Berry, to reproduce Beethoven's
ninth
symphony with such superoscillations requires a signal $e^{10^{19}}$ times
stronger than with conventional oscillations. A similar feature shows up
for black holes. In the specific case of a solar mass black hole
to reproduce a one-second thermal symphony would require a
signal $e^{10^{4}}$ stronger
in the superoscillating part of the evanescent hair at the horizon. In other
words, the incredible powerful signal is not to be found in the incoming
vacuum frequencies but can be accommodated by various evanescent quantum hairs
that any black hole
possesses at the horizon. Thus, my claim is that black hole hairs, a not
very exciting topic recently (due to the {\em no hair conjecture}),
may prove useful in solving the transplanckian paradox, if considered as
superoscillating reservoirs.

{\bf Acknowledgments:}
This work was partially supported by the CONACyT project 4868-E9406. I thank
Professor M.V. Berry for sending me his remarkable papers on
superoscillations.



\begin{thebibliography}{99}

\bibitem[*]{byline} Electronic-mail: rosu@ifug.ugto.mx

\bibitem{jac}
JACOBSON T.,
{\em Phys. Rev. D}, {\bf 44} (1991) 1731; ``Introduction to black hole
microscopy", hep-th/9510026, to appear in {\em Proc. of the first Mexican
school on gravitation and mathematical physics}, eds. A. Macias, T. Matos,
O. Obreg\'on and H. Quevedo (1996)~.

\bibitem{u}
UNRUH W.G., 
{\em Phys. Rev. D}, {\bf 51} (1995) 2827; {\em Phys. Rev. Lett.},
{\bf 46} (1981) 1351;
see also, BROUT R. et al., hep-th/9506124 and
CASHER A. et al., hep-th/9606106~.

\bibitem{tj}
JACOBSON T., {\em Phys. Rev. D}, {\bf 53} (1996) 7082~.

\bibitem{aapv}
AHARONOV Y., ANANDAN J., POPESCU S., and VAIDMAN L., {\em Phys. Rev. Lett.},
{\bf 64} (1990) 2965; See also,
VAIDMAN L., {\em Found. Phys.}, {\bf 21} (1991) 947;
SUTER D., {\em Phys. Rev. A}, {\bf 51} (1995) 45;
VAIDMAN L., {\em Phys. Rev. A}, {\bf 52} (1995) 4297~.

\bibitem{b}
BERRY MICHAEL, ``Faster than Fourier", in {\em Fundamental Problems in
Quantum Theory}, eds. J.A. Anandan and J. Safko (Singapore, World Scientific),
to appear, and
{\em J. Phys. A: Math. Gen.}, {\bf 27} (1994) L391~.

\bibitem{rosh}
ROSU H., work in progress~.



\end{thebibliography}
\end{document}